\documentclass[proceedings]{JHEP3}
\PrHEP{PrHEP hep2001}                   % Do not change this one
\conference{International Europhysics Conference on HEP}                % Do not change this one

\usepackage{epsfig}                   % please use epsfig for figures
\newcommand{\mydef}[2]{\def #1{\ifmmode{\rm #2}
          \else ${\rm #2}$\fi}}
\newcommand{\Res}[3]{\ifmmode{#1\:_{\footnotesize -#2}^{\footnotesize +#3}\mbox{\rule{0cm}{2.5ex}}} \else
$#1\:_{\footnotesize -#2}^{\footnotesize +#3}$\rule{0cm}{2.5ex}\fi}
\mydef{\Pgpp}{\pi^+}
\mydef{\PDm}{D^-}
\newcommand{\etal}{\mbox{$et$ $al.$}}
\def\aDz{\ifmmode{\mathrm{\overline{D}}^{\: 0}}\else ${\mathrm{\overline{D}}^{\: 0}}$\fi}
\def\Dz{\ifmmode{\mathrm{D}^{0}}\else ${\mathrm{D}^{0}}$\fi}
\def\gDz{\Dz}
\def\tilde{$\sim$}
\def\aDstz{\ifmmode{\mathrm{\overline{D}}^{\:\ast 0}}\else ${\mathrm{\overline{D}}^{\:\ast 0}}$\fi}
\def\Dstz{\ifmmode{\mathrm{D}^{\ast 0}}\else ${\mathrm{D}^{\ast 0}}$\fi}
\def\Bz{\ifmmode{{\mathrm{B}^0}}\else ${\mathrm{B}^0}$\fi}
\def\aBz{\ifmmode{\mathrm{\overline{B}}^0}\else ${\mathrm{\overline{B}}^0}$\fi}

\mydef{\degree}{^\circ}
\mydef\BR{{\cal B}}

\mydef{\ACP}{{\cal A}_{CP}}
\mydef{\Dtokpi}{\gDz\rightarrow K^- \pi^+}
\mydef{\Dtokpipi}{\gDz\rightarrow K^- \pi^+ \gpz}
\mydef{\Dtokpipipi}{\gDz\rightarrow K^- \pi^+\pi^+\pi^-}
\mydef{\mbc}{M_{bc}}
\mydef{\mbcc}{\mbc}
\mydef{\deltae}{\Delta E}
\mydef{\FD}{{\cal F}_D}
\mydef{\ctst}{\cos\theta_{B-Hel.}}
\mydef{\Btodpi}{\aBz\rightarrow\Dz\gpz}
\mydef{\Btodstpi}{\aBz\rightarrow\Dstz\gpz}
\mydef{\BBar}{B\overline{B}}
\mydef{\Dsttodpi}{\Dstz\rightarrow\Dz\gpz}
\mydef{\Dsttodg}{\Dstz\rightarrow\Dz\gamma}
\mydef{\Navfit}{<N_{fitted}>}
\mydef{\Nexp}{N_{Expected}}
\mydef{\RMSfit}{RMS_{fit}}
\mydef{\Significance}{N/\sigma(N)}
\mydef{\Nsig}{N_{Sig}}
\mydef{\Nbb}{N_{B\overline{B}}}
\mydef{\Ncont}{N_{Cont}}
\mydef{\Like}{L}
\mydef{\DDstz}{D^{(*)0}}
\mydef{\gpp}{\pi^+}
\mydef{\gpz}{\pi^0}
\mydef{\gpm}{\pi^-}
\mydef{\gppm}{\pi^\pm}
\mydef{\lepton}{\ell}
\mydef{\Btoddstpi}{\aBz\rightarrow\DDstz\gpz}
\mydef{\to}{\rightarrow}
\mydef{\Btosg}{b\to s \gamma}
\mydef{\btosg}{b\to s \gamma}
\mydef{\Btokll}{B\to K^{(*)} {\lepton}^+{\lepton}^-}
\mydef{\kshort}{K_s^0}
\newcommand{\CKM}[1]{\ifmmode{\rm V_{#1}}\else ${\rm V_{#1}}$\fi}
\mydef{\Btodkshort}{B^+\to D^{*+} \kshort}
\mydef{\BtodfpiMinus}{B^-\to D^{*0} \gpp\gpm\gpm\gpz}
\mydef{\BtodfpiZero}{\aBz\to D^{*0} \gpp\gpp\gpm\gpm}
\def\bsgResult{(3.03$\pm$0.40$\pm$0.26) \kreuz 10$^{-4}$}
\def\bsgacp{$(-0.079\pm0.108\pm0.022)(1.0\pm0.030)$}
\def\kllResult{1.7 \kreuz 10$^{-6}$}
\def\kstllResult{3.3 \kreuz 10$^{-6}$}
\def\kllaverageResult{1.5 \kreuz 10$^{-6}$}
\def\bdksResult{9.5 \kreuz 10$^{-5}$}
\def\bdfpiZeroResult{(0.30$\pm$0.07$\pm$0.06)\%}

\mydef{\geta}{\eta}
\mydef{\Jpsi}{J/\Psi}
\mydef{\myarrow}{{\rightarrow}}
\mydef{\kreuz}{\times}
\mydef{\yfours}{\Upsilon(4S)}
\mydef{\Ebeam}{E_{beam}}
\mydef{\Dtokpi}{\gDz\myarrow K^- \pi^+}
\mydef{\Dtokpipi}{\gDz\myarrow K^- \pi^+ \gpz}
\mydef{\Dtokpipipi}{\gDz\myarrow K^- \pi^+\pi^+\pi^-}
\mydef{\Dtokthreepi}{\gDz\myarrow K^-(3\pi)^\pm}
\mydef{\Kpi}{K^- \pi^+}
\mydef{\Kpipi}{K^- \pi^+ \gpz}
\mydef{\Kpipipi}{K^- \pi^+\pi^+\pi^-}
\mydef{\mbc}{M_{B}}
\mydef{\mbcc}{\mbc}
\mydef{\deltae}{\Delta E}
\mydef{\FD}{{\cal F}_D}
\mydef{\ctst}{\cos\theta_{B Hel.}}
\mydef{\Btodpi}{\aBz\myarrow\Dz\gpz}
\mydef{\Btodstpi}{\aBz\myarrow\Dstz\gpz}
\mydef{\BBar}{B\overline{B}}
\mydef{\Dsttodpi}{\Dstz\myarrow\Dz\gpz}
\mydef{\Dsttodg}{\Dstz\myarrow\Dz\gamma}
\mydef{\Navfit}{\overline{N_{fit}}}
\mydef{\Nexp}{N_{{\footnotesize Expect.}}}
\mydef{\RMSfit}{{\small RMS}}
\mydef{\Significance}{\frac{N}{\sigma_N}}
\mydef{\Nsig}{N_{Sig}}
\mydef{\Nbb}{N_{B\overline{B}}}
\mydef{\Ncont}{N_{Cont}}
\mydef{\Like}{{\cal L}}
\mydef{\DDstz}{D^{(*)0}}
\mydef{\Btoddstpi}{\aBz\myarrow\DDstz\gpz}
\mydef{\Efficiency}{\varepsilon}
\mydef{\rhm}{\rho^-}
\mydef{\Btoddstrho}{B^-\myarrow D^{(*)0}\rho^-}
\mydef{\Btodrho}{B^-\myarrow D^{0}\rho^-}
\mydef{\Btodstrho}{B^-\myarrow D^{*0}\rho^-}

\def\dsignalregion{$-0.05$ $<$ \deltae\ $<$ $0.05$ GeV, $5.275$ $<$ \mbc\
$<$ $5.285$ GeV}

\mydef{\BRdpi}{2.6\pm0.3\pm0.6}
\mydef{\BRdstpi}{2.0\pm0.5\pm0.7}

\def\cosdiold{0.90$\pm$0.09}
\def\cosdistarold{0.91$\pm$0.08}

\title{Rare and Hadronic B decays with CLEO}

\author{\speaker{Eckhard von Toerne}                       % One and only one
        for the CLEO Collaboration\\  % author must be
        Address: The Ohio State University, Columbus OH 43210\\                                          % inserted as
        E-mail: \email{evt@mps.ohio-state.edu}}                       % speaker
\abstract{Based on the CLEO II and II.5 data sets 
CLEO has observed several new rare and hadronic B decays
and also updated the \btosg\ measurement.}

\begin{document}
Rare and hadronic B meson decays provide an excellent experimental field 
to test the Standard Model of particle physics.
While effects like CP-violation have been observed in hadronic B decays $B
\to \Jpsi \kshort$,
the extraction of CKM-matrix elements and phases might still
prove difficult 
since many decay channels have to be measured and effects like final state interactions, 
re-scattering and interference between dominant
and suppressed decay amplitudes have to be understood \cite{heavy_flavor,browder_honscheid}. 
This makes it necessary to study extensively all rare and hadronic B
decays to gain full understanding of the dynamics in B decays.

The results presented here are based on the CLEO II and II.5 data sets.
The CLEO detector is located at the
CESR ${\rm e^+e^-}$ collider in 
Ithaca, NY. An integrated luminosity 
of 9.1 fb$^{-1}$ was collected on the \yfours\
resonance and 4.3 fb$^{-1}$ \tilde 60 MeV below the resonance to study the
continuum background from ${\rm e^+e^-\rightarrow q\bar{q}}$.
The kinematics of the \yfours\ decay, in which
two B mesons with equal masses are produced,
allow us to define two sensitive variables: 
the {Beam-constrained mass} ${\rm \mbc = \sqrt{E^2_{\textstyle
beam} - P^2_{B}}}$ and the 
{Energy difference} ${\rm \deltae = E_{B} - \Ebeam }$, where ${\rm
E_B}$ and ${\rm P_B}$ are the measured 
energy and momentum of the B candidate
and ${\rm E_{beam}}$ is the beam energy.
\section{${\boldmath\Btosg}$}
Many of the rare B decay modes can be enhanced by non-Standard Model 
contributions.
The most prominent example might be \btosg. This
flavor changing neutral current does not occur
as a first-order Standard Model process but is allowed via penguin
diagrams.
Non-Standard Model decays, where the W$^-$ in the penguin 
loop is replaced by a
charged Higgs or Chargino, would alter the branching fraction.

The \btosg\ measurement presented here has increased statistical significance
and updates a previous CLEO result \cite{old_bsg}.  The
model-dependence of 
the analysis has also been reduced by extending the
signal window from E$_\gamma$=2.2-2.7 GeV to 2.0-2.7 GeV.
\subsection{Extraction of the ${\boldmath\Btosg}$ signal}
\begin{figure}[htb]
\centerline{\epsfig{file=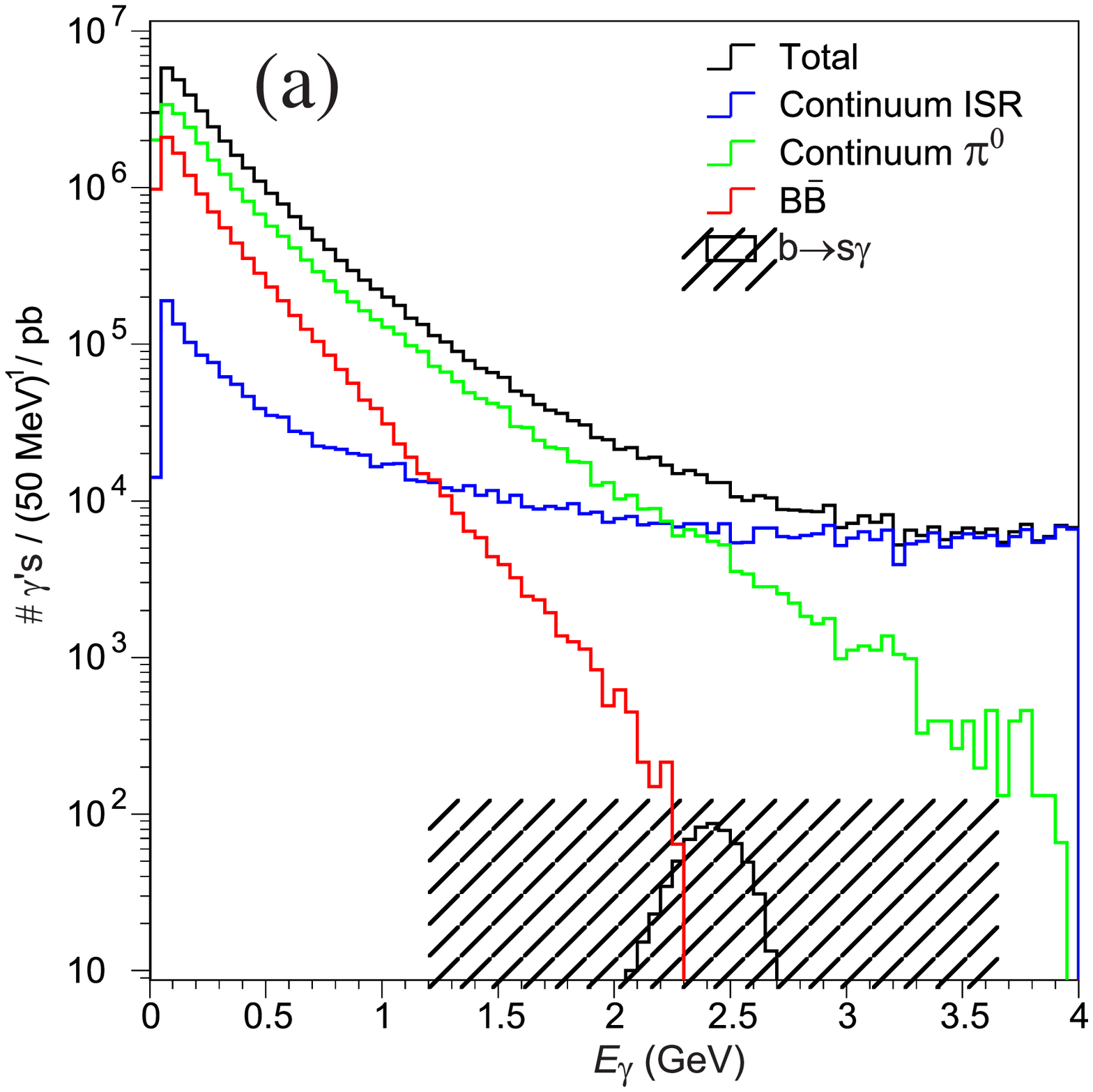,width=8cm}\epsfig{file=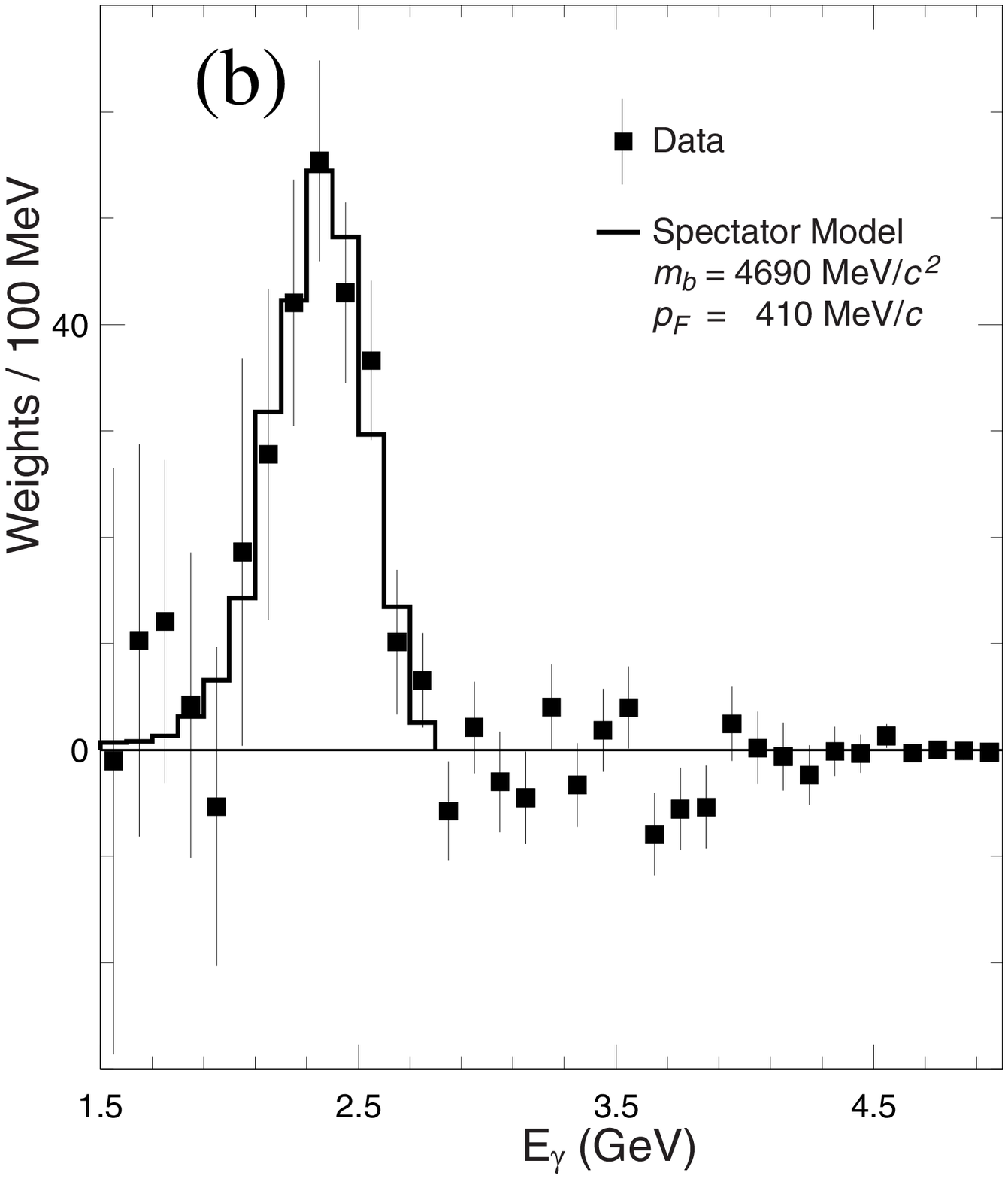,width=7cm,height=8cm}}
\begin{center}
\parbox{145mm}{\caption[ ]{\label{bsg.eps}\it (a) Signal and
background levels for \Btosg. (b) Photon energy spectrum after
background subtraction; a clear signal peak is observed at about
half the B mass. The spectator model fit is shown as the histogram.}}
\end{center}
\vspace*{-0.5cm}
\end{figure}

Even considering the inclusive sample, \btosg\ is a rare decay. 
\gpz\ or ${\rm\eta}$ mesons produce a large
background of real photons in addition to events with ISR.
Fig.~\ref{bsg.eps}-(a) demonstrates the signal and background levels on a
logarithmic scale.
The dominant background comes from continuum events. \BBar\ events
contribute with a much softer photon spectrum. 

Continuum ISR events are excluded by a cut on the polar angle
$|\cos\vartheta_\gamma|<0.7$.
Photons are vetoed if they can be combined with other photons 
to form a \gpz\ or \geta.
Continuum background is further suppressed by collecting event
information that can separate the continuum from \BBar, using
event shape variables and pseudo-reconstruction of hadronic B
decays. 
In the first pseudo-reconstruction method, B candidates are
formed from the photon candidate and additional kaon and pion tracks.
The second method uses identified leptons to tag semi-leptonic decays
of the companion B. Both
methods allow us to measure the flavor of the B, which will be used
in the CP-asymmetry measurement.

All the candidate's quantities are combined into one variable using a 
neural net. The neural net output is used as a weight, which peaks at +1
for \btosg\ events. The weighted
energy spectra for on- and off-resonance data (scaled to luminosity)
are subtracted. The weighting increases
the statistical significance of the on-resonance signal. 
The \BBar\ background is subtracted in two steps.
First we subtract the measured spectra of photons from
\gpz\ and \geta\ decay. The remaining background 
has contributions from B decays with successive 
electro-magnetic decays like
radiative \Jpsi\ or ${\rm \omega}$ decays. Only the tail of the
photon energy spectrum reaches the signal region. 
The remaining background
is estimated with a detailed Monte Carlo simulation and is
also subtracted.
The photon energy spectrum after subtraction 
is shown in Fig.~\ref{bsg.eps}-b. A signal peak is prominent at $E_\gamma\approx 1/2\: \mbc$.
The sidebands of the peak are consistent with zero. 
The preliminary branching fraction is \BR(\btosg)=\bsgResult\ and
differs slightly from our final
result \cite{bsgfinal}.
Both results are consistent with Standard Model calculations
\cite{bsg_sm1,bsg_sm2}. 

The E$_\gamma$ spectrum is fitted using a parameterization of the 
spectator model \cite{spectator_model} with the b-quark mass and the
fermi momentum of the b as free parameters.
The fit results are also shown
in Fig.~\ref{bsg.eps}-b. The calculation of these parameters 
can be utilized to reduce the uncertainty of \CKM{cb} and \CKM{ub}
measurements in semi-leptonic B decays \cite{vcb}.

The CP-asymmetry in \btosg\ is defined as ${\rm\ACP = (\Gamma(N_b)
-\Gamma(N_{\bar{b}}))/(\Gamma(N_b) +\Gamma(N_{\bar{b}}))}$.
We find no significant asymmetry: \ACP =\bsgacp\ \cite{bsgacp}. 
The first error is statistical,
the second is the additive systematic error, and the multiplicative 
systematic error comes 
from the systematic uncertainties on the tagging efficiency and
background subtraction.
We also derive 90\% C.L. limits on \ACP: $-0.27$ $<$ {\cal
A}$_{CP}$ $<$ $+0.10$.
\section{Searches for ${\boldmath\Btokll}$ and ${\boldmath\Btodkshort}$}
We searched for \Btokll, where \lepton\ is either an electron or a
muon. In decays with a ${\rm K^*}$ we impose a cut on the
${\lepton}^+{\lepton}^-$ invariant mass of
$m_{{\lepton}^+{\lepton}^-}>$ $0.5$ GeV to remove the contributions
from the virtual photon pole. We derive upper limits
of \BR(B \to K $\lepton^+ \lepton^-$) $<$ \kllResult,
\BR(B \to K$^*_{(892)} \lepton^+ \lepton^-_{m_{\lepton\lepton}>0.5\:GeV}$) $<$ 
\kstllResult \cite{kll}.
We obtain a combined limit of \BR(B \to K$^{(*)}\lepton^+ \lepton^-$)
$<$ \kllaverageResult\ at 90\% C.L., which is only 50\% higher than
the Standard Model prediction \cite{kll_sm}.

CLEO has also searched for the decay \Btodkshort\ \cite{bdkshort}.
This decay proceeds through an annihilation diagram that is
proportional to \CKM{ub}. 
We observe zero candidates while expecting
$0.29\pm0.05$ background events and derive an upper limit of \BR(B$^+$\to
D$^{*+}$\kshort) $<$ \bdksResult\ at 90\% C.L.
\section{Factorization tests in ${\boldmath\rm B\to D^{(*)}\;4\pi}$}
In a recent publication \cite{rhopr} CLEO has studied the decay 
${\rm B\to D^{(*)} \pi^-\pi^-\pi^+\pi^0}$. 
%and observed B\to D$^{(*)}\rho^{'-}$, $\rho^{'-}\to \omega\pi^-$.
The four-pion mass spectrum can be compared to the spectrum 
in tau decays ${\rm \tau^-\rightarrow\gpm\gpm\gpp\gpz\nu}$. 
A study by Ligeti, Luke and Wise \cite{ligeti_wise} demonstrated 
good agreement 
between B and tau decay data over the accessible mass range of
$0.6$--$1.7$ GeV.
Their comparison tested the factorization
hypothesis as a function of the four-pion mass. 
The validity of this test might be limited because
of additional contributions\footnote{E.g. B$\rightarrow (4\pi)^-$ decays
in which not all pions come from the virtual W$^-$.} to ${\rm B\rightarrow D^*
(4\pi)^-}$, that are not possible in tau decays. 
These additional 
contributions can be measured in a related decay mode \BtodfpiZero. 
%${\rm B\to}$${\rm D^{(*)} \pi^-\pi^-\pi^+\pi^+}$
This mode is suppressed
since the net charge of the pions is zero and the pions cannot come
from the W$^-$ decay alone. CLEO has observed this decay for the first
time with a branching fraction of \BR(\BtodfpiZero)=\bdfpiZeroResult \cite{bdfpizero}.
In the four-pion mass range $0.6$--$1.7$ GeV, 
the \BtodfpiZero\ contribution is consistent with zero.
This supports the validity of the 
factorization test by Ligeti \etal
\section{Color-suppressed decays ${\boldmath\Btoddstpi}$}
The decay \Btoddstpi\ proceeds predominantly through the internal
spectator diagram. 
%shown in Fig.~\ref{feynman.eps}. 
This decay mode is color-suppressed,  
since the color of the quark-pair originating from the W 
decay must match the color of the other quark pair.
The observation of \Btoddstpi\ completes the measurement of ${\rm
D^{(*)}\pi}$ final states and allows us to extract the strong phase difference
between isospin 1/2 and 3/2 amplitudes \cite{heavy_flavor,rosner99}. 
%\begin{figure}[b]
%\vspace*{-0.5cm}
%\centerline{\epsfig{file=feynman.eps,width=8cm}}
%\vspace*{-0.3cm}
%\begin{center}
%\parbox{145mm}{\caption[ ]{\label{feynman.eps}Diagram for the
%color-suppressed decay \Btoddstpi.}}
%\end{center}
%\end{figure}

B mesons are reconstructed by selecting high-momentum 
D$^{(*)0}$ and \gpz\ mesons.
The kinematic resolution of \gpz\ and D$^{(*)0}$ candidates
is improved by mass-constrained kinematic fits.
We accept B candidates with \mbc\ above 5.24 GeV and $|\deltae|$ $<$
300 MeV.
For each candidate we calculate the sphericity vectors of the B
daughter particles and of the rest of the event. 
We require the cosine of the angle between these two vectors to be within -0.8 and 0.8.
%The distribution of this angle is strongly peaked at $\pm$1 for
%continuum background and is nearly flat for \BBar\ events.

The signal yield is obtained from an
unbinned, extended maximum likelihood fit. The free parameters 
of the fit are
the number of signal events, background from B decays, and 
from continuum e$^+$e$^-$ annihilation.
Four variables are used as input to the maximum likelihood fit:
the beam-constrained mass \mbc, the energy difference \deltae,
the Fisher Discriminant \FD, which is a combination of event shape
variables, and the cosine of the decay angle of
the B \ctst, defined as the   
angle between the D$^{(*)}$ momentum and the B flight direction 
calculated in the B rest frame.
The likelihood of the B candidate is the sum of
probabilities for the signal and two background hypotheses with
relative weights maximizing the likelihood.
%The probability of a particular hypothesis is the product of
%probability density functions (PDFs) for each of the input variables.
%The PDF shapes are determined from off-resonance CLEO data
%(continuum) and from high-statistics Monte-Carlo (MC) samples 
%(signal and \BBar).
Monte Carlo experiments are performed to test the fitting procedure
and to obtain the relation between fit yield and signal branching
fractions. 

\begin{figure}[htb]
\centerline{\epsfig{file=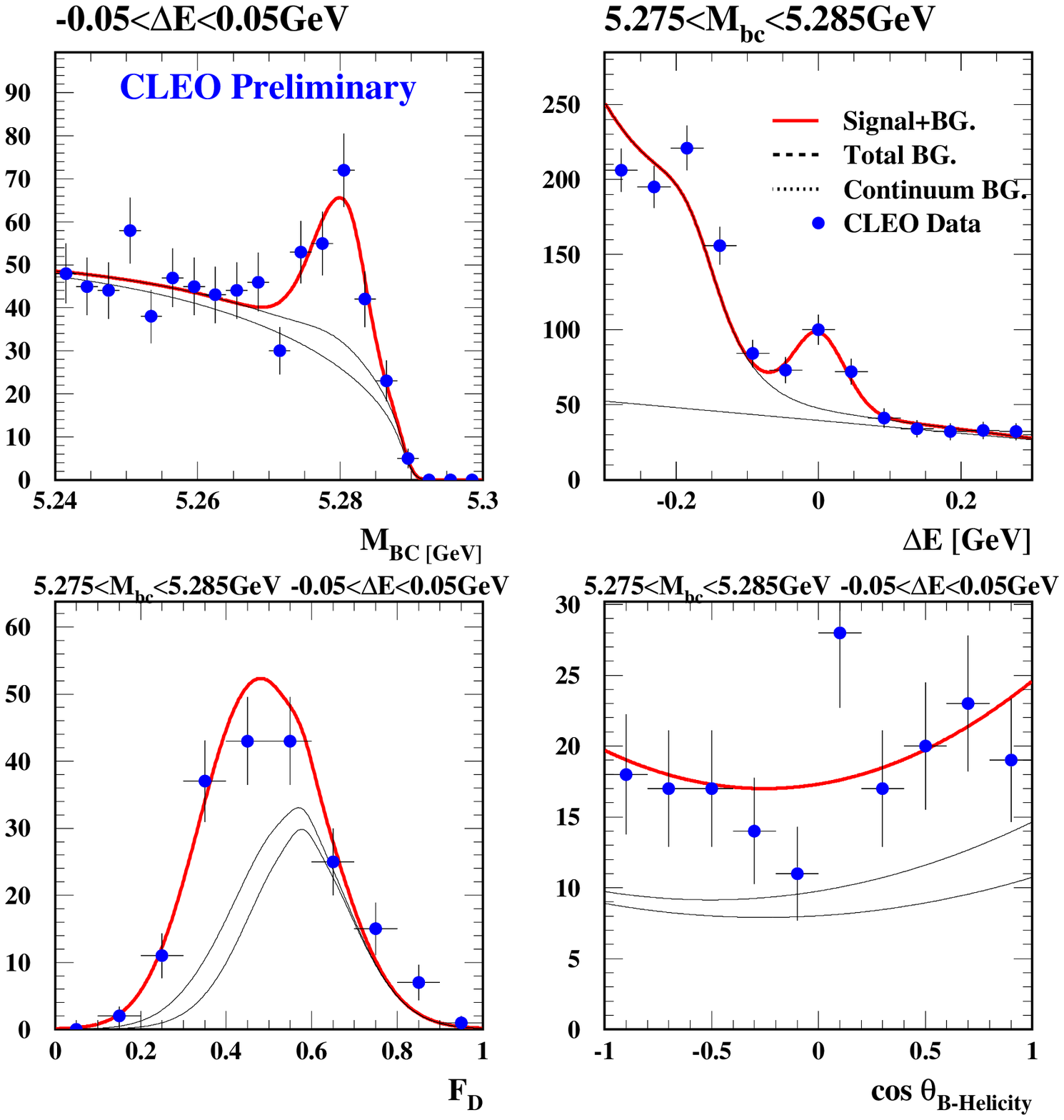,width=8.5cm}\hspace*{-1.0cm}\epsfig{file=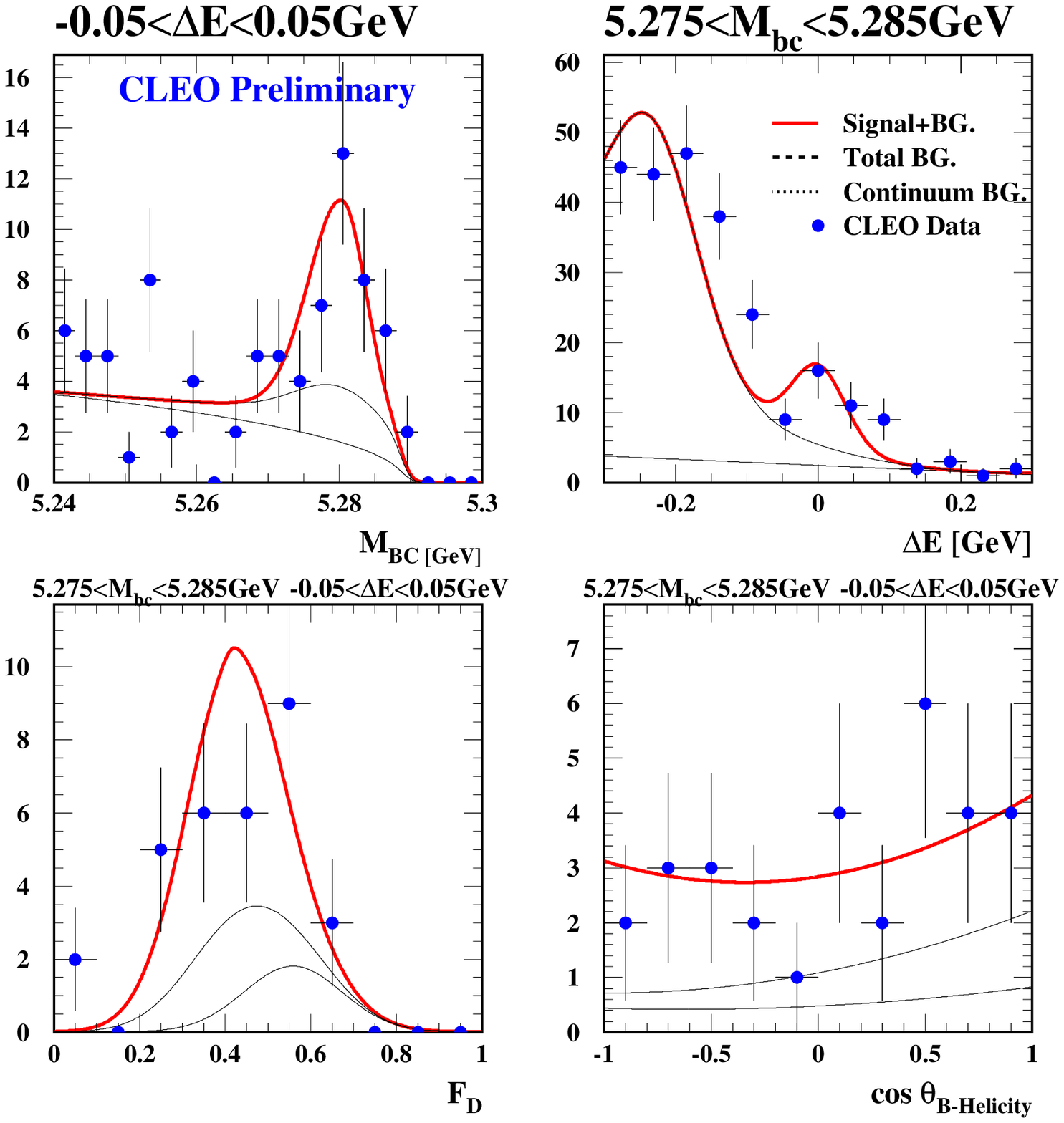,width=8.5cm}}
\begin{center}
\parbox{145mm}{\caption[ ]{\label{signalregion_overlay_bdpi0.eps}\it
Fit
input variable distributions. Left side \Btodpi, right side
\Btodstpi. 
The results of the unbinned, extended
maximum likelihood fit are shown as the 
full line. The dotted line represents the fitted 
continuum and the dashed line is the fit result for the sum of \BBar\ 
and continuum background. 
To enhance the signal for display purposes, the fit results are projected 
into the \mbc-\deltae\ signal region \dsignalregion.}}
\end{center}
\vspace*{-0.3cm}\end{figure}

We observe both \Btoddstpi\ decay modes with
preliminary branching fractions: 
$\BR(\Btodpi)$ = $(\BRdpi)\kreuz 10^{-4}$ and
$\BR(\Btodstpi)$ = $(\BRdstpi)\kreuz 10^{-4}$,
which differ slightly from our final results \cite{bdpi0_final}.
%Our result for \Btodpi\ is higher than the previous CLEO upper limit 
%\cite{bdpi0_cleo2}. 
%We ascribe this disagreement
%partly to a statistical fluctuation and 
%partly to the background description of the old CLEO
%publication.
Fig.~\ref{signalregion_overlay_bdpi0.eps} demonstrates 
the significance of our results which is above 4 sigma for each
decay mode.
Our measurements allow us to calculate the strong phase difference
between isospin 1/2 and 3/2 amplitudes. We obtain
$\cos \delta_{I,D}$=\cosdiold\ and $\cos \delta_{I,D*}$=\cosdistarold, 
for ${\rm D\pi}$ and ${\rm D^*\pi}$, respectively.
We quote the cosine of the angle because the errors on the cosine 
are gaussian to a good approximation. 
The error distribution of the angle $\delta_I$ is highly asymmetric
and non-gaussian. A detailed discussion can be
found in a publication by Neubert and Petrov
\cite{neubert_petrov_color_suppr} which is based on
our preliminary results
and preliminary results obtained by the BELLE Collaboration \cite{belle_eps01}.
%Models of hadronic B decay \cite{heavy_flavor} have successfully
%described experimental results using two phenomenological parameters,
%$a_1$ and $a_2$, that characterize non-factorizable contributions. Both are
%believed to be process-dependent but so far experimental data have
%been consistent with universal values for $a_1$ and $a_2$.
%Recent work by Beneke, Buchalla, Neubert and Sachrajda \cite{neubert_beneke} 
%has shown that $a_1$ is only slightly process-dependent.
%Comparing our result to two-body B decays to charmonium the
%process dependence of $a_2$ is favored \cite{neubert_petrov_color_suppr}.
%\section{Conclusion}
%\section{...}
% \EPSFIGURE{filename.eps}                       % if you need
% {Text of the caption.\label{figlabel}}         % to put a figure

% \TABLE{\begin{tabular}...
%       ....
%       \end{tabular}%
%       \caption{Text of the caption             % If you need to put
%                of the table.\label{tablabel}}} % a table


\begin{thebibliography}{99}
\bibitem{heavy_flavor}
M.~Neubert and B.~Stech in {\it Heavy Flavors}, 
edited by A.J.~Buras and M.~Lindner (World Scientific, Singapore, 
2nd edition 1998).
\bibitem{browder_honscheid}
T.~E.~Browder, K.~Honscheid and D.~Pedrini,
Ann.~Rev.~Nucl.~Part.~Sci.~{\bf 46}, 395 (1996).
\bibitem{old_bsg} 
CLEO Collaboration, M.~S.~Alam \etal, Phys.~Rev.~Lett.~ {\bf 74} 2885 (1995).
\bibitem{bsgfinal} 
Final result: \BR(\btosg)=$(3.21\pm0.43\pm0.27^{+0.18}_{-0.10}) \kreuz 10^{-4}$ \\
CLEO Collaboration, S.~Chen \etal, hep-ex/0108032, subm.~to Phys.~Rev.~Lett.
\bibitem{bsg_sm1} K.~Chetyrkin, M.~Misiak, and M.~M\"{u}nz,
Phys.~Lett. {\bf B400}206 (1997).
\bibitem{bsg_sm2} P.~Gambino and M.~Misiak, hep-ph/0104034.
\bibitem{spectator_model} A.~Ali and C.~Greub, Phys.~Lett.~{\bf B259}
182 (1991). 
\bibitem{vcb} D.~Miller, same proceedings; 
and D.~Cronin-Hennessy \etal\ (CLEO), hep-ex/0108033. 
\bibitem{bsgacp} CLEO Collaboration, T.~E.~Coan \etal,
Phys.~Rev.~Lett.~{\bf 86} 5661 (2001).
\bibitem{kll} CLEO Collaboration, S.~Anderson \etal, 
Phys.~Rev.~Lett.~{\bf 87} 181803 (2001).
\bibitem{kll_sm} A.~Ali \etal, Phys.~Rev.~D {\bf 61} 074024 (2000).
\bibitem{bdkshort} CLEO Collaboration, A.~Gritsan \etal, Phys.~Rev.~D
{\bf 64} 077501, (2001).
\bibitem{rhopr} CLEO Collaboration, J.~P.~Alexander \etal, 
Phys.~Rev.~D {\bf 64} 092001 (2001).
\bibitem{ligeti_wise} Z.~Ligeti, M.~Luke, and M.~B.~Wise, 
Phys.~Lett.~{\bf B507} 142 (2001).
\bibitem{bdfpizero} CLEO Collaboration, K.~W.~Edwards \etal,
hep-ex/0105071, subm.~to Phys.~Rev.~Lett.
%\bibitem{bdpi0_cleo2} CLEO Collaboration, {B. Nemati \etal}, 
%Phys.~Rev.~D {\bf 57}, 5363 (1998).
\bibitem{rosner99} 
J.L.~Rosner, Phys.~Rev.~D {\bf 60}, 074029 (1999).
%\bibitem{PDG} D.E.~Groom \etal, (Particle Data Group), Eur.~Phys.~Jour.~{\bf C15}, 1 (2000) and 2001 partial update for edition 2002 (URL: http://pdg.lbl.gov).
\bibitem{bdpi0_final} 
CLEO Collaboration, T.~E.~Coan \etal, hep-ex/0110055, subm.~to 
Phys.~Rev.~Lett. F{\small inal} R{\small esults:} \BR(\Btodpi)=$({2.74}_{\footnotesize -0.32}^{\footnotesize
+0.36}\pm0.55)\kreuz10^{-4}$, \BR(\Btodstpi)=$({2.20}_{\footnotesize
-0.52}^{\footnotesize +0.59}\pm0.79)\kreuz10^{-4}$
\bibitem{neubert_petrov_color_suppr} M.~Neubert and A.~Petrov, 
Phys.~Lett.~{\bf B519}, 50 (2001).
\bibitem{belle_eps01} R-S.~Lu (same proceedings) and hep-ex/0107048;
Final results by the BELLE Collaboration: hep-ex/0109021, subm.~to Phys.~Rev.~Lett.
%\bibitem{neubert_beneke} M. Beneke \etal, Nucl.Phys.~{\bf B591} (2000) 313.
\end{thebibliography}
\end{document}